\documentclass[manuscript]{aastex}

\slugcomment{Not to appear in Nonlearned J., 45.}

\shorttitle{The Saturation Levels For White-Light Flares of The Flare Stars}
\shortauthors{Dal and Evren}

\begin{document}


\title{Saturation Levels For White-Light Flares of The Flare Stars:\\
Variation of Minimum Flare Durations For The Saturation}


\author{H. A. Dal and S. Evren}
\affil{Department Of Astronomy and Space Sciences, University of Ege, \\
Bornova, 35100 ~\.{I}zmir, Turkey}

\email{ali.dal@ege.edu.tr}



\begin{abstract}

In this study, considering the results obtained from the models and from the statistical analyses of the obtained parameters, flare activity levels and flare characteristics of five UV Ceti stars will be discussed. We also present the parameters of unpublished flares detected in two years of observations of V1005 Ori. The parameters of the U-band flares detected in several seasons of observations of \object{AD Leo}, \object{EV Lac}, \object{EQ Peg}, \object{V1054 Oph} and \object{V1005 Ori} are compared among themselves. The flare frequencies calculated for all program stars and maximum energy levels of the flares are compared and it is discussed which is the most correct parameter as an indicator of the flare activity levels. Using \textit{One Phase Exponential Association} function, the distributions of the flare equivalent durations versus the flare total durations were modelled for each programme star. We used the \textit{Independent Samples t-Test} in the statistical analyses of the parameters obtained from the models. The results revealed some properties of the flare processes running on the surfaces of UV Ceti type stars: 1) Flare energies can not be higher than a specific value and it is no matter how long flare total duration is. This must be a saturation level for white-light flares occurring in flare processes observed in U band. Thus, in the first time it is shown that white-light flares have a saturation in the energy range. 2) The \textit{span values}, which is the difference between the equivalent durations of flares with the shortest and longest total durations, are almost equal for each star. 3) Another important result to model the white-light flares is that the \textit{half-life values}, minimum flare durations for the saturation, increases towards the later spectral types. 4) Both maximum total durations and maximum rise times computed from the observed flares decrease towards the later spectral types among the UV Ceti stars. According to the maximum energy levels obtained from the models, both \object{EV Lac} and \object{EQ Peg} are more active than other three program stars, while \object{AD Leo} is the most active flare star according to the flare frequencies.

\end{abstract}





\section{Introduction}

UV Ceti type stars are mostly pre-main sequence stars coming to the ZAMS or the stars at the ZAMS. Most of the red dwarfs in open clusters and associations are flare stars \citep{Mir90, Pig90}. The number of flare stars in the clusters decreases with the increasing age of clusters. This is an acceptable case according to Skumanich Law \citep{Sku72, Mar92, Pet91, Sta91}. Higher rotation rate cause high flare activity level and this causes higher mass loss by flare bursts. The studies in literature indicate that the mass loss rate of the Sun is about $2x10^{-14}$ $M_{\odot}$ in per year \citep{Ger05}. However, this value for UV Ceti type stars can reach to the value of $10^{-10}$ $M_{\odot}$ in per year because of the flare activity. The high level of mass loss for these stars can explain that they lose 98 $\%$ of their angular momentum in the course of main sequence \citep{Mir90}.

Magnetic flare activity causing high mass loss has not been properly explained. The average of the energies of classical flares of the Sun is $10^{26}$ - $10^{27}$ ergs. It is about $10^{30}$ - $10^{31}$ ergs for two ribbon flares, which are known as the hardest flares on the Sun \citep{Ger05, Ben10}. When flares of the chromospherically active stars, known as RS CVn stars, are examined, it is seen that flare energies of these stars are about $10^{31}$ ergs \citep{Hai87}. On the other hand, observations lasting about 50 years show that flare energies of UV Ceti stars change from $10^{28}$ ergs to $10^{34}$ ergs \citep{Ger05}. Moreover, when flare stars in the young clusters such as Pleiades and Orion associations are considered, flare energies can reach the value of $10^{36}$ ergs \citep{Ger83}.

As it can be seen from literature, there are distinctive differences between flare energies of different type stars. Nevertheless, although there are clear differences between the Sun and UV Ceti type stars in point of mass loss and flare energies, flare activity of dMe stars is modelled by basing on the processes of Solar Flare Event. In this case, it is accepted that source of the energy in these events is the magnetic reconnection process \citep{Ger05, Hud97}. To understand properly all process of flare event for dMe stars, first of all, it could be useful to find out all similarities and differences of flare light curves by examining star-to-star. For instance, looking over flare energy spectra can be helpful as the first step. This distribution can show how flare energies range located from one star to other. In this respect, both the distributions and the levels of the flare energy spectra of UV Ceti type stars have been looked over in many studies such as \citet{Ger72, Lac76, Wal81, Ger83, Pet84} and, \citet{Mav86} might be given as the examples. For instance, flare energy spectra have been studied by \citet{Ger72} for \object{AD Leo}, \object{EV Lac}, UV Cet and YZ CMi. In another study made by \citet{Ger83}, flare energy spectra for lots of stars in galactic field have been compared with flare energy spectra of some stars from Pleiades cluster and Orion association. As it is seen from Figure 2b given by \citet{Ger83}, the flare energy spectra for the stars of Orion association are located at very high levels from others. Then Pleiades stars are located below the stars of Orion association. Finally, the stars from galactic field are located below them. This distinctive separation among flare energies of these stars from different origin shows that something is different in the flare process on these stars. It seems that the differences between the levels of the flare energy spectra are due to different ages. On the other hand, there is some separation among the stars located in galactic field, too.

All these differences among the different stars and star groups might be due to the saturation level of white-light flares detected from UV Ceti type stars. It is seen that some parameters of magnetic activity can reach the saturation \citep{Ger05, Sku86, Vil83, Vil86, Doy96a, Doy96b}. White-light flares are detected in some large active regions such as compact and two-ribbon flares occurring on the surface of the Sun \citep{Rod90, Ben10}. We expect that the energies or the flare equivalent durations can reach the saturation. The analyses of some large flare data sets can demonstrate this expectation. If a saturation level can be found, it will be a guidance to model the white-light flares. However, the data sets using in the analyses are important. The data sets are combined from the parameters derived with the same method from the flares detected with the same optic system. Otherwise, some artificial variations and differences can occur between the sets. To avoid this problem, we used a large data sets, which were combined from the parameters derived with the same method and the same optic system.

In the second place, flare frequencies can be another useful parameter to understand the flare process. In some studies such as \citet{Pet83, Ish91, Let97}, flare frequencies have been examined. Two flare frequencies have been generally calculated in these studies. As it is seen from literature, flare energies and frequencies are varying from star to star. To understand all flare process, the reason(s) of these variations should be find out. In this point, it must be answered to whether these differences are because of some physical parameters such as mass and age of stars, or are directly due to different flare processes?

In this study, we compared flare parameters among five UV Ceti type stars given in Table 1. The physical parameters (such as mass, radius and the distance) taken from \citet{Ger99} are given in Table 1. According to spatial velocities taken from \citet{Mon01}, it is seen that \object{AD Leo} is a red dwarf member of Castor Moving Group, whose age is about 200 million years. In the same way, according to the spatial velocities, \object{EV Lac} is seen as a member of 300 million years old Ursa Major Group, which is known as the Sirius Supercluster. \object{V1005 Ori}, whose first flare light curve was obtained by \citet{Sha74}, is a member of 35 million years old IC 2391 Supercluster \citep{Mon01}. According to another study made by \citet{Vee74}, these three stars seem to belong to the young disk population of the galaxy, too. \object{EQ Peg}, whose distance is 6.58 pc, is classified as a metal-rich star and it belongs to the young disk population of the galaxy. On the other hand, \object{V1054 Oph} is a triple-lined spectroscopic system as known Gliese 644 (=Wolf 630+Wolf629ABab). The distance of the system is 6.5 pc. \object{V1054 Oph} is classified as a metal-rich star and a member of the old disk population of the galaxy \citep{Vee74, Fle95}. The flare stars, whose flare parameters are compared in this study, are quite young stars except \object{V1054 Oph}. The masses were derived for each components of Wolf 629ABab by \citet{Maz01}. They showed that the masses are 0.41 $M_{\odot}$ for Wolf 629A, 0.336 $M_{\odot}$ for Wolf 629Ba and 0.304 $M_{\odot}$ for Wolf 629Bb. In addition, \citet{Maz01} demonstrated that the age of the system is about 5 Gyr.

\section{Observations and Analyses}

\subsection{Observations}

The observations were acquired with the High-Speed Three Channel Photometer attached to the 48 cm Cassegrain type telescope at Ege University Observatory. Using a tracking star set in the second channel of photometer, the observations of the variable star were continued in standard Johnson U band with the exposure time between 2 and 10 seconds. The basic parameters of all the program stars (such as standard V magnitudes and B-V colours) are given in Table 2. The parameters given in Table 2 were taken from \citet{Dal10} for \object{AD Leo}, \object{EV Lac}, \object{EQ Peg} and \object{V1054 Oph}. Although the program and comparison stars are so close on the plane of sky, differential extinction corrections were applied. The extinction coefficients were obtained from the observations of the comparison stars on each night. Moreover, the comparison stars were observed with the standard stars in their vicinity and the arrested differential magnitudes, in the sense variable minus comparison, were transformed to the standard system using procedures outlined by \citet{Har62}. The standard stars are listed in the catalogues of \citet{Lan83, Lan92}. Heliocentric corrections were applied to the times of the observations. The standard deviation of each observation acquired in standard Johnson U band is about $0^{m}.15$ on each night. Observational reports of all the program stars are given in Table 3. The differential magnitudes in the sense comparison minus check stars were carefully checked for each night. The comparison and check stars were found to be constant in brightness during the period of observation. Equivalent durations and energies of all the flares are computed from photoelectric observations using Equation (1) and (2) taken from \citet{Ger72}:

\begin{center}
\begin{equation}
P = \int[(I_{flare}-I_{0})/I_{0}] dt
\end{equation}
\end{center}
where, $I_{0}$ is the flux of star in the observing band during the star in the quiet state; $I$ is the intensity at the moment of flare.

\begin{center}
\begin{equation}
E = P \times L
\end{equation}
\end{center}
where $E$ is the energy. $P$ is the flare equivalent duration in the observing band. $L$ is the intensity in the observing band during the star in the quiet state. For each observed flare, HJD of flare maximum moment, flare rise and decay time, flare amplitude, flare equivalent duration and their energy have been calculated. In the calculations, the quiescent level, in which there are no flare and other variability, has been accepted as the basic level of light curve in nightly light curves. All the parameters have been computed by considering this level. Some flares have several peaks. If a flare has several peaks, maximum point of the flare has been accepted as the peak, which is close to beginning and the highest one. Instead of the flare energy, flare equivalent duration has been used in the comparison. This is because of the luminosity term in Equation (2). The luminosities of stars with different spectral types have large differences. Although the equivalent durations of two flares detected from two stars in different spectral types are the same, calculated energies of these flares are different due to different luminosities of these spectral types. Therefore, we could not use these flare energies in the same analysis. On the other hand, flare equivalent duration depends just on power of the flare. Another reason of using equivalent duration is that the given distances of the same star in different studies are quite different. These differences cause the calculated luminosities become different.

In Figure 1, a fast flare sample detected in observations of \object{V1005 Ori} on 6 January 2005 is seen. According to the rule described described by \citep{Dal10}, this flare is classified as a fast flare. Consecutive two flares detected on 10 January 2005 are shown in Figure 2. These flares are classified as a slow flare in respect to the same rule. The consecutive flare samples of \object{V1005 Ori} are shown in Figure 3; these flares were detected on 29 December 2005.

Using the method described by \citet{Dal10}, all parameters are calculated for each flare detected in the observations of \object{V1005 Ori}. The calculated parameters for 41 U-band flares are given in Table 4. The columns of table are the date of observation, HJD of flare maximum, flare rise time, flare decay time, equivalent duration, flare amplitude and flare type, respectively.

\subsection{Analyses}

\object{V1005 Ori} flare data were combined with the data set including 321 U band flares detected from other stars (\object{AD Leo}, \object{EV Lac}, \object{EQ Peg} and \object{V1054 Oph}). The parameters of 321 U band flares detected from these stars were presented by \citet{Dal10}. Using this large data set including total 362 U band flares, program stars are compared with each other in the analyses to find out whether there is any differences between their flare activity behaviours. In this point, as it is mentioned above, the data using in the analyses are important. The data must be combined from the parameters determined by the same method. In addition, the flares must be observed with the same optic systems. There are large data sets including U band flares in the literature such as \citet{Mof74, Ish91}. On the other hand, the methods used to determine the parameters of the detected flares are not the same in these studies and there are some differences between some optic system. This is why we used the data obtained in this study and the data presented by \citet{Dal10}. All the parameters in these data sets were determined with the same method and all the flares were detected with the same optic system.

Instead of flare energies, flare equivalent durations were used for all statistical analyses. This is because of the luminosity term in Equation (2). The luminosities of stars from different spectral types have great differences. Although the equivalent durations of two flares obtained from two stars in different spectral types are the same, calculated energies of these flares are different due to different luminosities of these spectral types. Therefore, we could not use these flare energies in the same analyses. On the other hand, flare equivalent duration depends just on flare power. Another reason of using equivalent duration is that the given distances of the same star in different studies are quite different. Therefore, the calculated luminosities become different because of these different distances.

When the distribution of flare equivalent durations in logarithmic scales versus flare total durations is examined, it is seen that flare equivalent duration is below a limit value and it is no matter how long flare total duration is. In addition, when we compare stars among themselves, it is clear that limit value of energy is different for each star.

To model the distributions the best function was searched. Using the SPSS V17.0 \citep{Gre99} and GrahpPad Prism V5.02 \citep{Daw04} softwares, the regression calculations showed that the best fit is \textit{One Phase Exponential Association} (hereafter \textit{OPEA}) for the distributions of flare equivalent durations. The \textit{OPEA} function given by Equation (3) \citep{Mot07, Spa87} is a special exponential function, which has a plateau part, as it is seen in the distributions of flare equivalent durations. Using the Least-Squares Method, for each star the distributions were modelled by the \textit{OPEA} function. All the distributions and their models with 95 $\%$ confidence intervals are shown in Figure 4.

\begin{center}
\begin{equation}
y~=~y_{0}~+~(Plateau~-~y_{0})~\times~(1~-~e^{-k~\times~x})
\end{equation}
\end{center}
where $y$ is the flare equivalent duration in logarithmic scales, while $x$ is flare total duration. $y_{0}$ is flare equivalent duration in logarithmic scales for the least total duration. In other words, $y_{0}$ is the least equivalent duration occurring in a flare for a star. The value of $y_{0}$ depends on the brightness of the target and the sensitivity of the optic systems. The value of \textit{Plateau} is the upper limit for the equivalent duration, which can be occurred in a flare for a star. This parameter can be identified as a saturation level for flare activity observed in U band. According to Equation (2), the value of \textit{Plateau} depends only on the energy of flares occurring on the star. According to the definition of the \textit{OPEA} function, the parameter $k$ in Equation (3) is a constant depending on the $x$ values.

The parameters derived from the \textit{OPEA} models are given for all the stars in Table 5. Star name, B-V index, \textit{Plateau}, $y_{0}$ and $k$ values are listed in the table, respectively. The \textit{span value} and \textit{half-life value} are given in the last two columns. B-V indexes are also found in this study. The \textit{span value} is the difference between the values of \textit{Plateau} and $y_{0}$. The \textit{half-life value} is half of the first $x$ values, where the model starts to give the \textit{Plateau} values for a star. In other words, it is half of the flare total duration, where flares with the highest energy start to seen. The \textit{half-life value} is minimum value of the flare total duration, which is needed for the saturation level.

Considering the flare in the plateau phases of the \textit{OPEA} models, all programme stars were compared in respect of maximum flare equivalent durations. \textit{Independent Samples t-Test} (hereafter \textit{t-Test}), which is a statistical analysis method \citep{Wal03, Daw04, Mot07}, was used for comparisons. Using this test, the mean equivalent durations were computed for each star and the mean values were compared among themselves. The results are given in Table 6. Thus, the values of \textit{Plateau} were tested with another analysis.

Although the mean values computed by \textit{t-Test} are expected to be close to the \textit{Plateau} values of the \textit{OPEA} models for all stars, it is clear that there can be some difference between these two values. This is because the \textit{OPEA} models depend on all distribution from the beginning of $x$ values to the end, while the mean values computed by \textit{t-Test} depend only on equivalent durations of flares in the \textit{Plateau} phases.

The variations of \textit{Plateau} values, which are listed in Table 5, and the mean values of equivalent durations, which are given in Table 6, are plotted versus the B-V of each star and shown in Figure 5. As it is expected, both parameters exhibit the same variations with a little difference in their levels. In the figure, the mean equivalent durations and the \textit{Plateau} values decrease with increasing B-V index for three stars, namely \object{V1005 Ori}, \object{AD Leo} and \object{V1054 Oph}. On the other hand, these two parameters are dramatically higher for two stars, \object{EV Lac} and \object{EQ Peg}, which are the reddest stars among five stars. In the figure, decreasing of first three values is shown by linear fits for both parameters. The variation of \textit{Plateau} values indicates that the saturation level of flare activity can change. Moreover, the saturation levels of the reddest stars among five stars is absolutely higher than the levels of other stars.

The variations of other parameters derived from the \textit{OPEA} models are shown versus the B-V index in Figure 6. The variation of the parameter $y_{0}$ is seen in panel (a) of this figure. Although the parameter of $y_{0}$ depends on both the brightness of target and the sensitivity of the observing system in a general manner due to the standard deviations of observations, here $y_{0}$ parameter exhibits a dramatic increase for two stars, which are located towards the reddest edge. It means that the flare energy for the least total duration of flares is higher for the stars, which are located towards the reddest edge of M type. In panel (b), the variation of the \textit{span values}, which is the difference between $y_{0}$ and \textit{Plateau} values obtained from model curves, is seen versus B-V index. According to the models, the \textit{span value} shows no important variation versus B-V index. In panel (c), the variation of \textit{half-life values} is seen versus B-V index. The regression calculations indicated that the polynomial function is the best fit, which is shown by dotted line in the figure, for this variation of the value. As it is seen from the panel, the \textit{half-life values} increase towards the reddest M stars in respect of analyses of these five stars.

In addition, when 362 flares observed from the five stars are examined, it is seen that the longest flare total durations are varying from star to star. As it is seen from Figure 4, the total duration of the longest flare is 2940 seconds for \object{EV Lac} and 3180 seconds for \object{EQ Peg}, while it is 5236 seconds for \object{V1005 Ori} and it reaches to 4164 seconds for \object{AD Leo}. The longest flare total duration is about 3270 seconds for \object{V1054 Oph}. As it is seen in Figure 7, the observed maximum flare total durations decrease towards the later spectral types. The maximum flare rise time was also computed for each star. For \object{V1005 Ori}, \object{AD Leo}, \object{V1054 Oph}, \object{EV Lac} and \object{EQ Peg}, the maximum rise times are 2036, 1212, 1460, 840 and 1230 seconds, respectively. The variation of all these times is also shown in Figure 7.

Moreover, two different flare frequencies of stars were computed for each season to examine flare activity levels. These flare frequencies are identified by Equation (4) and (5) taken from \citet{Ish91}.

\begin{center}
\begin{equation}
N_{1}~=~\Sigma n_{f}~/~\Sigma T_{t}
\end{equation}
\end{center}

\begin{center}
\begin{equation}
N_{2}~=~\Sigma P_{u}~/~\Sigma T_{t}
\end{equation}
\end{center}
where $\Sigma T_{t}$ is the total observing durations, $\Sigma n_{f}$ is the total number of flares obtained in a season ($\Sigma n_{f}$) and $\Sigma P_{u}$ is the total equivalent duration obtained from all flares detected in that observing season ($\Sigma P_{u}$). $N_{1}$ and $N_{2}$ are the flare frequencies. All computed frequencies are listed in Table 7. According to the results, the higher flare frequencies are seen in \object{AD Leo}.

\section{Results and Discussion}

The flare energy, which is expressed by Equation (2) given by \citet{Ger72}, has been generally used to examine the level of flare activity in lots of studies in the literature. The studies of some authors such as \citet{Maz01}, \citet{Lac76}, \citet{Wal81}, \citet{Ger83}, \citet{Pet84} and \citet{Mav86} can be given as examples. The luminosity parameter ($L$) takes place in the expression of the energy ($E$), which has been based by all these and other studies. The luminosity ($L$) is different for each star. Although there are little differences among the masses of M dwarfs, the luminosities of two M dwarfs, whose masses are so close to each other, can be dramatically different from each others due to their places in the Hertzsprung-Russell diagram. This means that the computed energies of flares are very different from each other, even if the light variations of the flares occurring on these two stars are the same. Because of this, the equivalent durations ($P$) were used in the analyses instead of energy ($E$) in this study. If there is a difference in the equivalent durations of the flares, it is also seen in the energies in the same way.

In the analyses, the distributions of flare equivalent durations versus flare total duration were modelled by the \textit{OPEA} function expressed by Equation (3) for all stars in the programme. When the models are compared, it is seen that there are some differences among the stars. As it is seen in Figure 4f and Figure 5a, the \textit{Plateau} parameter, which gives the maximum equivalent duration level for flares on a star, is changing from one star to the other. It is seen in Figure 4f that the distributions of the equivalent durations in logarithmic scale versus flare total duration for \object{EV Lac} and \object{EQ Peg} are different from other three stars. The maximum equivalent durations seen in these two stars are as high as 0.5 times in logarithmic scales. This difference in logarithmic scales is equal to 683 times difference in energies. This means that, for example, the energy of an \object{EV Lac} flare is 683 times higher than the energy of an \object{AD Leo} flares in average generally. In addition, the energy of a flare occurring on \object{AD Leo} is never higher than the energy of an \object{EV Lac} flare, no matter how long the total duration of \object{AD Leo} flare is. This result is confirmed by another analysis, which is a statistical analysis method, \textit{t-Test}. We used the flares, whose equivalent durations in logarithmic scales are located in the plateau phases in the \textit{OPEA} models, in \textit{t-Test}. In this point, the aim is to compare the equivalent durations of flares, whose energies are independent from lengths of their total duration. The results of \textit{t-Test} analyses are shown in Figure 5b. As it is seen from this figure, the mean averages of equivalent durations computed by \textit{t-Test} are close to the \textit{Plateau} values derived from the \textit{OPEA} models. The mean averages of maximum equivalent durations for flares of \object{EV Lac} and \object{EQ Peg} are distinctively higher than the averages computed from other three stars. On the other hand, the mean averages of equivalent durations are relatively different for each star. 

As it is mentioned above, some parameters in the chromospheric magnetic activity can reach the saturation \citep{Ger05, Sku86, Vil83, Vil86, Doy96a, Doy96b}. In the case of the white-light flare, we expect that the energies or the flare equivalent durations can reach the saturation, because these white-light flares are detected in some large active regions such as compact and two-ribbon flares occurring on the surface of the Sun \citep{Rod90, Ben10}. Consequently, according to this approach, the \textit{Plateau} value must be a saturation level (or an indicator at least) for the white-light flares. In the analyses, we used the data obtained with the same method and the same optic system. In addition, we used the flare equivalent durations instead of the flare energies. Therefore, the derived \textit{Plateau} values depend just on the power of the white-light flares. Considering the \textit{Plateau} values, it is seen that the power of the flare has a limit for a star. The flare equivalent durations can not be higher than a value and it is no matter how long flare total durations. Instead of the flare duration, some other parameters, such as magnetic field flux and/or particle density in the volumes of the flare processes, must be more efficient to determine the power of the flares. Considering thermal and non-thermal flare events, both magnetic field flux and particle density in the volumes of the flare processes can be efficient.

\citet{Gur77, Gur88} developed a hypothesis, called Fast Electron Hypothesis. In this hypothesis, the source of the white-light flares on the surfaces of UV Ceti stars is non-thermal process, such as the spontaneous appearance of fast electrons on the surface of the flare stars. Considering this hypothesis, the particle density in the volumes of the flare processes must be more efficient to determine the power of the white-light flares, instead of magnetic field. \citet{Gur88} demonstrated that the inverse Compton effect, non-thermal interactions of infrared photons with fast electrons, causes some radiative losses. It is possible that the inverse Compton effect can be more efficient after a specific flare durations for a UV Ceti star, and this effect can limited the observed flare equivalent duration (and energy) of a detected flare. However, considering all the flare process, it should be noted that the particle density in the volumes depends on magnetic field flux in some respects. The source of theme is some particles accelerated by magnetic field \citep{Ben10, Ger05}. In addition, the magnetic field flux in the volumes is more efficient than the particle density for high energy patterns of the flare process, such as soft X-ray or radio intensities \citep{Ger05}.

On the contrary, \citet{Doy96a, Doy96b} suggested that the saturation in the active stars does not have to be related to the filling factor of magnetic structures on the stellar surfaces or the the dynamo mechanism under the surface, it can be related to some radiative losses in the chromosphere, where the temperature and density are increasing in the case of the fast rotation. Like this phenomenon, a case can occur in the chromosphere due to the flare process instead of the fast rotation, and this case causes that the \textit{Plateau} phase occurs in the distributions of flare equivalent durations versus flare total duration. On the other hand, the \textit{Plateau} phase can not be due to some radiative losses in the chromosphere with increasing of the temperature and density. This is because, \citet{Gri83} demonstrated the effects of the radiative losses in the chromosphere on the white-light photometry of the flares. According to \citet{Gri83}, the negative H opacity in the chromosphere causes the radiative losses, and these radiative losses seen as a pre-flare dips in the light curves of the white-light flares. In addition, when the results are considered, it is seen that the \textit{Plateau} values are varying from a star to next one. This indicates that some parameters, which cause the \textit{Plateau} in the distributions of flare equivalent durations, or their efficacies are changing star-to-star. Moreover, in the case of \object{EV Lac} and \object{EQ Peg}, there is a distinctive difference. The efficacies of these parameters (or the parameters themselves) must be dramatically changed for these stars.

In the future, the reason(s) of the \textit{Plateau} phases should be examined by synchronous observations in the radio, optic and X-ray regions of the spectrum. In these studies, we recommend that some tests should be made. When the energy of a white-light flare detected with optical photometry reaches the saturation level, it should be tested whether the energy reaches the saturation in the radio or X-ray observation, or not. If the energy reaches the saturation in the radio or X-ray observation, this indicates that the reason of the saturation is generally magnetic field. This is because, the energy source in the radio or X-ray is generally magnetic reconnection. If the energy does not reach the saturation in the synchronous optical photometry, it means that the particle density in the volumes is more efficient to determine the power of the flares in the optical part of the spectrum.

However, it is worthy to note, that known very week correlation of optical and radio flares at the UV Cet type variables makes such experiment hardly realized. Moreover, the attribution of flares in saturation regions to white-light flares seems to be weekly proved. On the Sun such flares are selected with spectroscopic observations when there is a strong continuum. On stars multicolour observations allow defining a phase when a black body radiation that is a continuum dominates \citep{Zhi07}.

When variation of the parameter $y_{0}$ is examined, as it is seen from Figure 6a, it is changing from star to star. Like the parameter \textit{Plateau}, $y_{0}$ parameters of \object{EV Lac} and \object{EQ Peg} are rather higher than other stars. $y_{0}$ parameters of other three stars are almost equal to each other. Actually, the parameter of $y_{0}$ depends on both the brightness of the target and the sensitivity of the observing system. On the other hand, considering that the brightness of all the targets are almost equal to each other and all of them were observed with the same system in almost the same time, the variation of $y_{0}$ parameters from star to star are close to the real behaviour.

The difference between \textit{Plateau} and $y_{0}$ parameters, which is derived from the distribution modelled by Equation (3), is listed in Table 5 as the parameter of \textit{span value}. Figure 6b shows the behaviour of this parameter versus B-V index. As it is seen, there is no regular variation in this parameter. It is important because it means that the difference between \textit{Plateau} and $y_{0}$ parameters is constant along all B-V indexes of M type in respect of five stars. The similarity of \textit{span values} for all programme stars shows that the difference between $y_{0}$ parameters of stars is exactly seen between \textit{Plateau} values of stars. This indicates that even if the conditions, in which the flares occur, and so energies of flares are changing, the difference between minimum and maximum energies is stable in flare mechanism.

At the same time, the variation of \textit{half-life values}, which is an indicator of the minimum flare total duration needed for maximum energy occurring, is seen in Figure 6c. According to this variation, the values of \textit{half-life value} increases towards higher B-V indexes. It means that the minimum total durations, which are needed for the flares emitting maximum energy in the flare mechanisms, increase towards the later spectral types among the M type stars. Consequently, this variation indicates that longer flare total durations are needed to reach the saturation level towards the later spectral types among M dwarfs. Like the variation of the \textit{Plateau} value versus B-V colour index, the variation of \textit{half-life values} must be considered to model the white-light flares detected from UV Ceti type stars.

When we consider maximum flare total durations seen among the flares for each star, the maximum flare total durations seen among the flares are 2940 seconds for \object{EV Lac} and 3180 seconds for \object{EQ Peg}. It is 5236 seconds for \object{V1005 Ori} and 4164 seconds for \object{AD Leo}, while it is about 3270 seconds for \object{V1054 Oph}. Maximum flare total durations seen in flares of UV Ceti stars decrease towards the later spectral types among the M stars as seen in Figure 7. Like flare total durations, the maximum rise times are 2036, 1212, 1460, 840 and 1230 seconds for \object{V1005 Ori}, \object{AD Leo}, \object{V1054 Oph}, \object{EV Lac} and \object{EQ Peg}, respectively. The variation of all these times is also shown in Figure 7. As it is seen from the figure, both maximum rise time and total duration decrease towards the later spectral types.

As a result, four important properties can be summarised for the flare processes occurring on the UV Ceti type stars: 1) Flare energies increase with flare total duration until a specific total duration values, and then the energies are constant no matter how long the flare total duration is. 2) The differences between minimum and maximum energies of flares are constant and the same for all stars. 3) The minimum total durations, which are needed for the flares emitting maximum energy, increase towards the later spectral types among the UV Ceti stars. 4) Maximum flare rise time and total durations decrease towards the later spectral types among the UV Ceti stars.

On the other hand, two flare frequencies expressed by Equation (4) and (5) are used to identify the flare activity levels in lots of studies in the literature. The researchers have used these frequencies to find out whether flare activity of UV Ceti stars exhibits any cyclic variation. Assuming that all flares occurring on the star are observed; $N_{1}$ is an indicator of flare number obtained in unite-time, as it is expressed by Equation (4). However, according to Equation (5), $N_{2}$ is an indicator of the mean equivalent duration average obtained in unite-time \citep{Ish91}. In brief, $N_{1}$ refers how many flares occur on a star, while $N_{2}$ refers how energetic these flare are.

In this study, both flare frequencies of $N_{1}$ and $N_{2}$ were computed season-to-season for each programme star. All of them are listed in Table 7. $N_{1}$ frequency of \object{AD Leo} is close to value of 1.0 in the season of 2005. It means that one flare occurred on \object{AD Leo} per hour at least. $N_{2}$ frequency of the star was computed as 0.086. Considering the values obtained for each season and each stars, this value of $N_{2}$ indicates that the flares occurring on \object{AD Leo} were powerful in the season of 2005. $N_{1}$ and $N_{2}$ frequencies are in agreement to each others in the season of 2005. Conversely, $N_{1}$ is 1.331, while $N_{2}$ is 0.012 in the season of 2006. $N_{1}$ is higher than the value of 1.0, and it can be accepted as the highest value of $N_{1}$, but $N_{2}$ is not high. Although lots of flares were able to occur on \object{AD Leo} in the season of 2006, their energies were not as high as expected values, as it was seen in the season of 2005. According to the value of $N_{2}$, flare activity level of \object{AD Leo} was not high. This is an argumentative case. Like the case of \object{AD Leo}, some similar cases are seen in other programme stars. In the literature, both $N_{1}$ and $N_{2}$ frequencies are accepted as an indicator of flare activity level in some studies. For example, \citet{Mav86} computed both $N_{1}$ and $N_{2}$. Examining distribution of these parameters versus time, they demonstrated that \object{EV Lac} had a flare activity cycle of 5 years.

According to expression given by Equation (5), $N_{2}$ depends on equivalent duration, in other words, energy. So, it is expected that $N_{2}$ can behave like the \textit{Plateau} or the mean average of the equivalent durations. $N_{2}$ frequencies of \object{EV Lac} and \object{EQ Peg} are expected to be higher than the same frequencies of the other stars. However, the frequencies of these two stars are almost the same. There is no clear difference between them.

According to all these results, the parameters of the \textit{Plateau} and the mean average of the equivalent durations seem to be useful to determine the flare activity levels. We assume that it is the most active star, whose \textit{Plateau} and mean parameters are the highest.

On the other hand, if the differences, which are seen between values of the \textit{Plateau} or the mean average of the equivalent durations, between the programme stars had been caused by the age of the stars, it was expected that \object{V1005 Ori} was the most active star among the others, according to \citet{Sku72}. This is because \object{V1005 Ori} is a member of \object{IC 2391} Supercluster, which is 30-35 million years old \citep{Mon01}. Considering that all other stars are almost in the same age apart from \object{V1054 Oph}, whose age is 5 Gyr, the difference can be caused by rapid rotation or binarity \citep{Vee74, Fle95, Mon01}. The equatorial rotational velocity of \object{EV Lac} is 4 $kms^{-1}$ and it is between 5 - 5.8 $kms^{-1}$ for \object{AD Leo} \citep{Mar92, Pet91}. Conversely, the equatorial rotational velocity of \object{V1005 Ori} is 29.6 $kms^{-1}$ \citep{Eke08}. Assuming that Skumanich Law can be acceptable for flare activity of UV Ceti stars, as it is in the case of chromospheric activity, the values of the \textit{Plateau} or the mean average of the equivalent durations must be higher for the stars, which are rapidly rotating. However, according to our results, this is not a common rule. Beside this, \object{EV Lac}, \object{AD Leo} and \object{V1005 Ori} are single stars. There is a visual companion of \object{EQ Peg}. Moreover, \object{V1054 Oph} is a system composed from six stars \citep{Pet91}. Therefore, if the binarity or multiplicity makes the flare activity levels increase; the most active flare star should be \object{V1054 Oph}.

We analysed the data obtained in three observing seasons from the observations of five stars. Finally we have already reached to some clear results. These results are derived from both the \textit{OPEA} models and the computed the flare frequencies. Obtained properties about flare processes occurring on UV Ceti type stars are important to understand the general flare process for stellar flare activity. In this respect, extending the B-V range of the programme stars, it is needed to obtain much more data, which are obtained from lots of different stars and their flare patrols spanning long years, in order to reach more reliable results.

\section*{Acknowledgments} The authors acknowledge generous allotments of observing time at Ege University Observatory. We wish to thank Dr. Hayal Boyac{\i}o\v{g}lu, who gave us important suggestions about statistical analyses. Prof. Dr. M. Can Akan gave us valuable suggestions, which improved the language of the manuscript, we wish to thank him. We also thank the referee for useful comments that have contributed to the improvement of the paper. We finally thank the Ege University Research Found Council for supporting this work through grant Nr. 2005/FEN/051.

\clearpage

\begin{figure}
\epsscale{.80}
\plotone{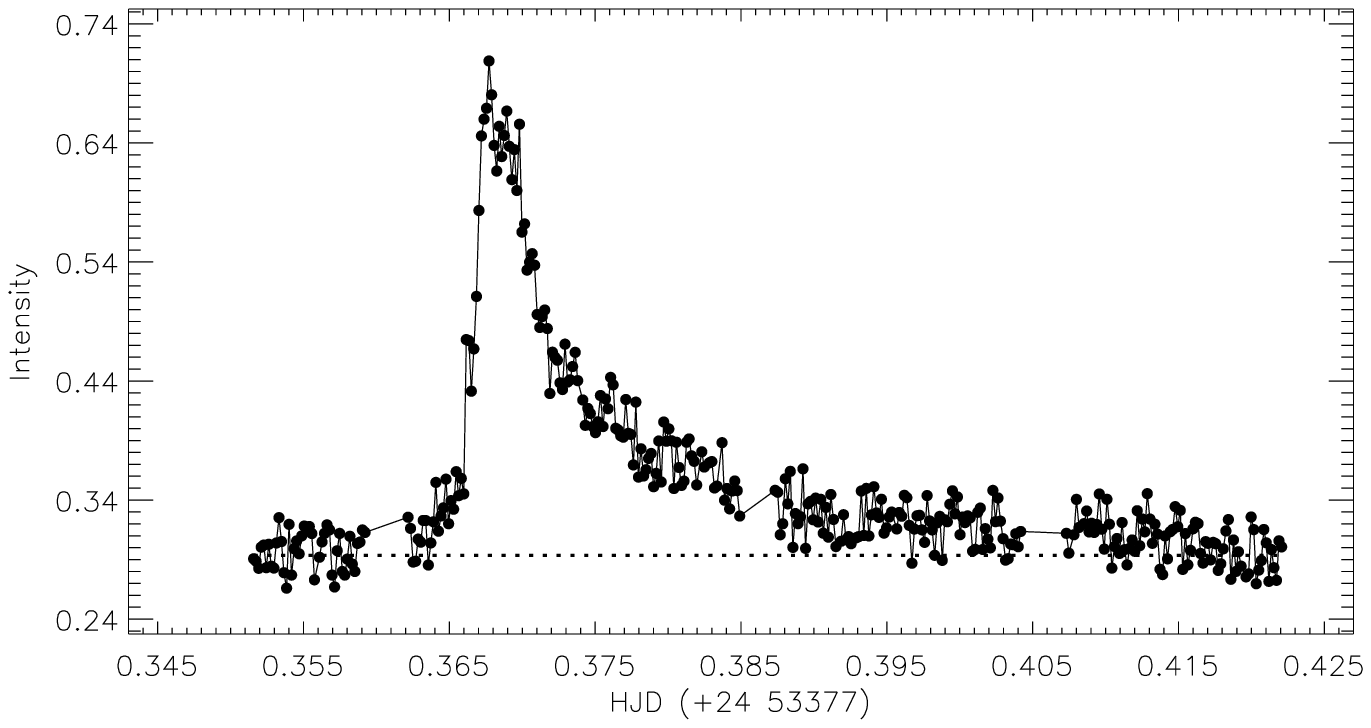}
\caption{A fast flare sample detected in U band observation of V1005 Ori on 6 January 2005. In the figure, filled circles show observations, while horizontal dashed line represents the level of quiescent state of the star in U band for the observing night.\label{fig1}}
\end{figure}

\begin{figure}
\epsscale{.80}
\plotone{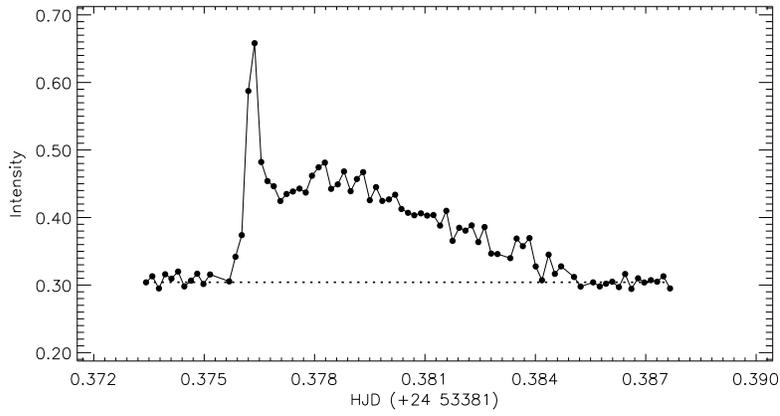}
\caption{A fast flare sample detected in U band observation of V1005 Ori on 10 January 2005. In the figure, all the symbols are the same as Figure 1.\label{fig2}}
\end{figure}

\begin{figure}
\epsscale{.80}
\plotone{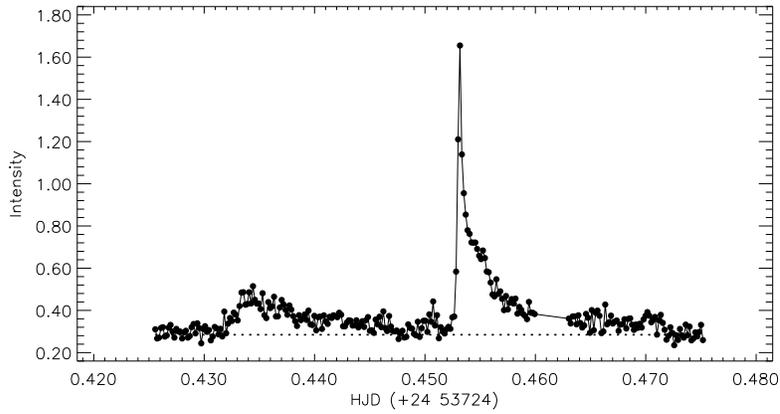}
\caption{A fast flare sample detected in U band observation of V1005 Ori on 29 December 2005. In the figure, all the symbols are the same as Figure 1.\label{fig3}}
\end{figure}

\begin{figure}
\epsscale{1.0}
\plotone{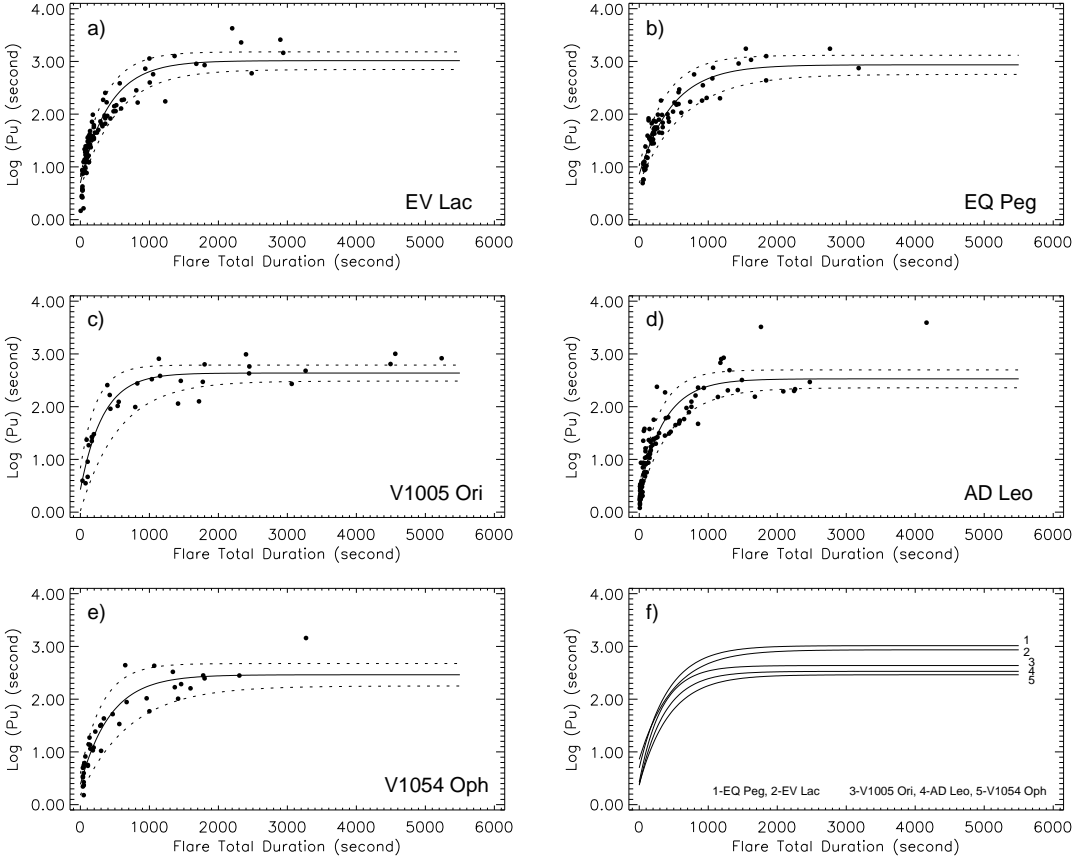}
\caption{The distributions of flare equivalent durations in logarithmic scales versus flare total durations for each programme star (a, b, c, d, e). Filled circles represent equivalent durations computed from observed flares. The lines represent the models identified with Equation (3) computed by using The Least-Squares Method. The dotted lines represent the 95 $\%$ confidence intervals for the models for each star. In panel (f), all models derived for each star are compared.\label{fig4}}
\end{figure}

\begin{figure}
\epsscale{.80}
\plotone{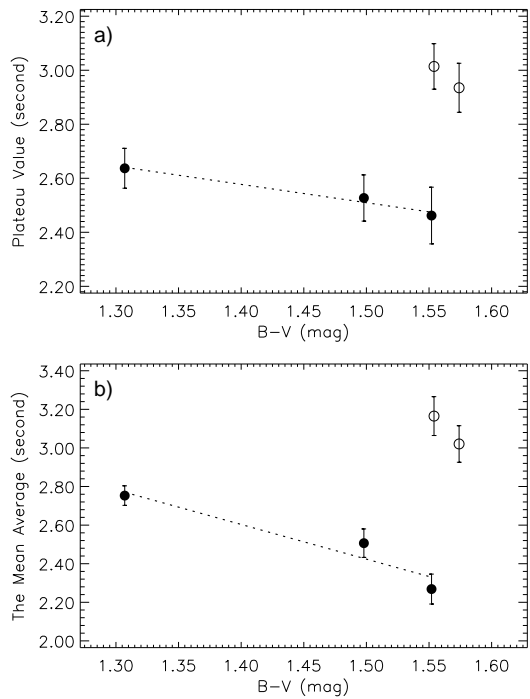}
\caption{The variations of both the \textit{Plateau} parameters (a) and the mean equivalent durations (b) are demonstrated versus B-V index of stars. The dotted lines represent the linear fits for the first three points, which are used to indicate decreasing of both values of the \textit{Plateau} and the mean equivalent durations.\label{fig5}}
\end{figure}

\begin{figure}
\epsscale{.90}
\plotone{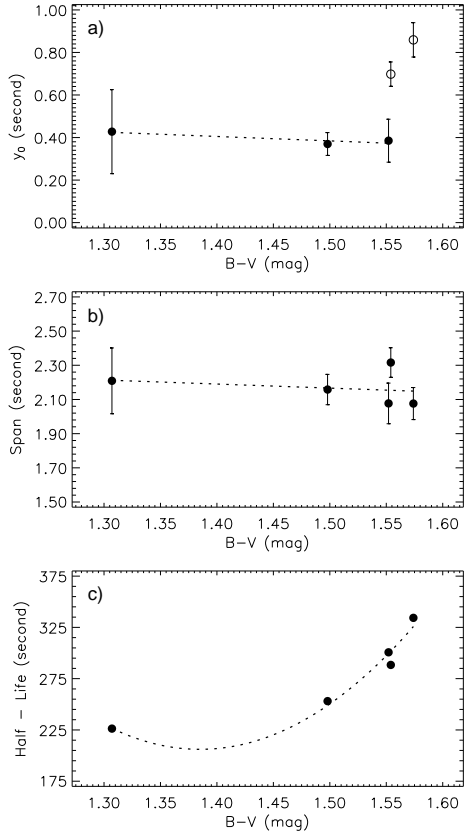}
\caption{The variations of both the parameter y0 (a) and the \textit{span value} (b) are shown versus B-V index of stars. In panel (c), the variation of the \textit{half-life values} is seen. The dotted line in panel (a) represents the linear fit for the first three points, as the same in Figure 5. In panel (b) it represents the linear fit for all points. It is the polynomial fit of all points in panel (c). All fits are only used for representation of variation ways.\label{fig6}}
\end{figure}

\begin{figure}
\epsscale{.90}
\plotone{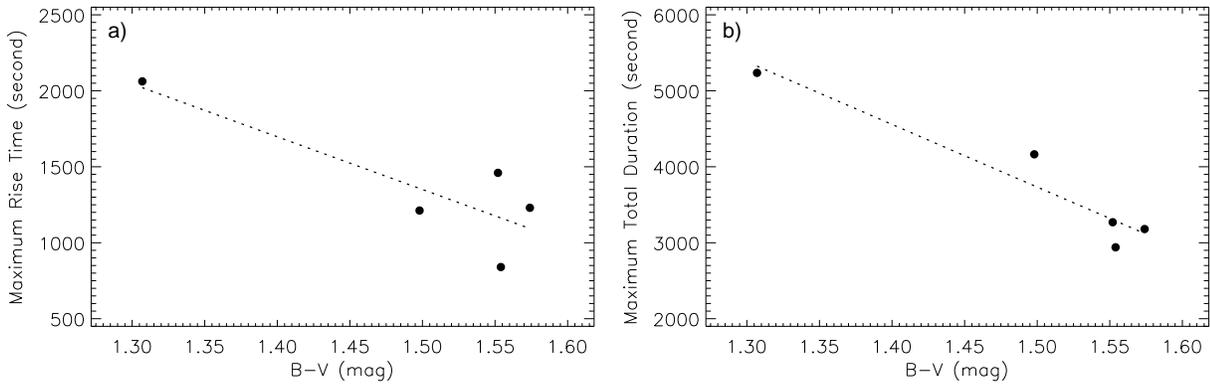}
\caption{The variations of maximum flare rise time (a) and maximum flare total durations (b) are seen versus B-V index. The dotted lines are the linear fit of them.\label{fig7}}
\end{figure}

\clearpage

\begin{table}
\begin{center}
\caption{Some physical parameters of program stars taken from \citet{Ger99}.\label{tbl-1}}
\begin{tabular}{lccc}
\tableline\tableline
\textbf{Star}	&	\textbf{Distance (pc)}	&	\textbf{Mass ($M_{\odot}$)}	&	\textbf{Radius ($R_{\odot}$)}	\\
\tableline 							
\textbf{AD Leo}	&	4.9	&	0.28	&	0.54	\\
\textbf{EV Lac} 	&	5	&	0.18	&	0.39	\\
\textbf{EQ Peg}	&	6.2	&	0.28	&	0.58	\\
\textbf{V1054 Oph}	&	5.7	&	0.42	&	0.76	\\
\textbf{V1005 Ori}	&	26.7	&	-	&	0.7	\\
\tableline
\end{tabular}
\end{center}
\end{table}

\begin{table}
\begin{center}
\caption{Basic parameters for the targets studied and their comparison (C1) and check (C2) stars. Columns list: Star name; standard V mag and B-V colours for quiet phase of them.\label{tbl-2}}
\begin{tabular}{lcc}
\tableline\tableline
\textbf{Star}	&	\textbf{V (mag)}	&	\textbf{B-V (mag)}	\\
\tableline 					
\textbf{V1005 Ori}	&	10.090	&	1.307	\\
C1= BD+01 870 	&	8.800	&	1.162	\\
C2= HD 31452	&	9.990	&	0.920	\\
\tableline 
\textbf{AD Leo} & 9.388 & 1.498 \\
C1 = HD 89772 & 8.967 & 1.246 \\
C2 = HD 89471 & 7.778 & 1.342 \\
\tableline 
\textbf{EV Lac} & 10.313 & 1.554 \\
C1 = HD 215576 & 9.227 & 1.197 \\
C2 = HD 215488 & 10.037 & 0.881 \\
\tableline 
\textbf{EQ Peg} & 10.170 & 1.574 \\
C1 = SAO 108666 & 9.598 & 0.745 \\
C2 = SAO 91312 & 9.050 & 1.040 \\
\tableline
\textbf{V1054 Oph} & 8.996 & 1.552 \\
C1 = HD 152678 & 7.976 & 1.549 \\
C2 = SAO 141448 & 9.978 & 0.805 \\
\tableline
\end{tabular}
\end{center}
\end{table}

\begin{table}
\begin{center}
\caption{Observational reports of the program star for each observing season.\label{tbl-3}}
\begin{tabular}{lccccc}
\tableline\tableline
\textbf{Star}	&	\textbf{Season}	&	\textbf{HJD Interval}	&	\textbf{Number}	&	\textbf{Observing}	&	\textbf{Flare}	\\
	&		&	\textbf{(+24 00000)}	&	\textbf{Of Night}	&	\textbf{Duration (h)}	&	\textbf{Number}	\\
\tableline 											
\textbf{V1005 Ori}	&	2004-2005	&	53353-53453	&	9	&	28.13	&	10	\\
	&	2005-2006	&	53673-53769	&	9	&	26.45	&	31	\\
\tableline
\end{tabular}
\end{center}
\end{table}

\rotate

\begin{deluxetable}{ccrrrccr}
\tabletypesize{\scriptsize}
\tablecolumns{9}
\tablewidth{0pc}
\tablecaption{All the calculated parameters of flares detected in the observations of V1005 Ori are listed. From the first column to the last, the date of observation, HJD of flare maximum moment, flare rise time, decay time, equivalent duration, flare energy (ergs), flare amplitude and flare type are given, respectively. In the last column, flare types are listed.}
\\
\tablehead{
\colhead{Observing}	&	\colhead{HJD Of Maximum}	&	\colhead{Flare Rise}	&	\colhead{Flare Decay}	&	\colhead{Equivalent}	&	\colhead{Energy}	&	\colhead{Amplitude}	&	\colhead{Flare}	\\
\colhead{Date}	&	\colhead{(+ 24 00000)}	&	\colhead{Time (s)}	&	\colhead{Time (s)}	&	\colhead{Duration (s)}	&	\colhead{(ergs)}	&	\colhead{(mag)}	&	\colhead{Type}}	\\
\\
\startdata														
13.12.2004	&	53353.42334	&	420	&	375	&	98	&	1.6799E+33	&	0.304	&	Slow	\\
13.12.2004	&	53353.43480	&	30	&	75	&	5	&	7.9712E+31	&	0.271	&	Slow	\\
13.12.2004	&	53353.44452	&	30	&	510	&	103	&	1.7617E+33	&	0.121	&	Fast	\\
06.01.2005	&	53377.31547	&	15	&	15	&	4	&	6.6193E+31	&	0.158	&	Fast	\\
06.01.2005	&	53377.36773	&	345	&	2055	&	980	&	1.6731E+34	&	0.919	&	Fast	\\
10.01.2005	&	53381.37637	&	60	&	765	&	276	&	4.7061E+33	&	0.834	&	Fast	\\
10.01.2005	&	53381.41977	&	450	&	1995	&	427	&	7.2863E+33	&	0.221	&	Fast	\\
12.01.2005	&	53383.26550	&	15	&	60	&	4	&	6.0373E+31	&	0.215	&	Fast	\\
12.02.2005	&	53414.34290	&	15	&	90	&	9	&	1.5447E+32	&	0.212	&	Fast	\\
12.02.2005	&	53414.35470	&	15	&	15	&	4	&	6.7371E+31	&	0.215	&	Slow	\\
29.10.2005	&	53673.50034	&	1450	&	3291	&	920	&	1.5697E+34	&	0.109	&	Fast	\\
29.10.2005	&	53673.56398	&	1661	&	2903	&	1005	&	1.7161E+34	&	0.100	&	Slow	\\
11.11.2005	&	53686.55961	&	2062	&	3174	&	827	&	1.4109E+34	&	0.189	&	Slow	\\
11.11.2005	&	53686.61910	&	1384	&	1065	&	577	&	9.8525E+33	&	0.383	&	Slow	\\
12.12.2005	&	53717.38316	&	750	&	1768	&	1824	&	3.1135E+34	&	0.309	&	Fast	\\
12.12.2005	&	53717.41844	&	636	&	1982	&	1621	&	2.7672E+34	&	0.694	&	Fast	\\
12.12.2005	&	53717.45388	&	750	&	1050	&	634	&	1.0814E+34	&	0.436	&	Slow	\\
12.12.2005	&	53717.47628	&	255	&	135	&	256	&	4.3697E+33	&	1.036	&	Slow	\\
19.12.2005	&	53724.34506	&	405	&	630	&	331	&	5.6530E+33	&	0.339	&	Slow	\\
19.12.2005	&	53724.40173	&	840	&	615	&	308	&	5.2619E+33	&	0.303	&	Slow	\\
19.12.2005	&	53724.41949	&	165	&	495	&	136	&	2.3156E+33	&	0.273	&	Fast	\\
19.12.2005	&	53724.42539	&	15	&	105	&	19	&	3.1694E+32	&	0.380	&	Fast	\\
19.12.2005	&	53724.42730	&	60	&	105	&	22	&	3.8074E+32	&	0.356	&	Slow	\\
19.12.2005	&	53724.43182	&	15	&	15	&	6	&	9.7737E+31	&	0.386	&	Fast	\\
19.12.2005	&	53724.43407	&	180	&	975	&	383	&	6.5315E+33	&	0.553	&	Fast	\\
19.12.2005	&	53724.44623	&	75	&	120	&	30	&	5.1338E+32	&	0.323	&	Slow	\\
19.12.2005	&	53724.45074	&	45	&	45	&	24	&	4.0302E+32	&	0.428	&	Slow	\\
19.12.2005	&	53724.45317	&	120	&	1017	&	810	&	1.3818E+34	&	1.889	&	Fast	\\
19.12.2005	&	53724.46633	&	30	&	405	&	91	&	1.5554E+33	&	0.412	&	Fast	\\
24.12.2005	&	53729.38460	&	1835	&	2660	&	643	&	1.0976E+34	&	0.073	&	Slow	\\
24.12.2005	&	53729.44269	&	1293	&	1971	&	479	&	8.1814E+33	&	0.150	&	Slow	\\
24.12.2005	&	53729.49801	&	30	&	135	&	27	&	4.5391E+32	&	0.319	&	Fast	\\
08.01.2006	&	53744.34648	&	136	&	289	&	166	&	2.8409E+33	&	0.662	&	Slow	\\
08.01.2006	&	53744.40960	&	1482	&	3472	&	2673	&	4.5624E+34	&	0.090	&	Fast	\\
27.01.2006	&	53763.30671	&	390	&	1384	&	296	&	5.0442E+33	&	0.267	&	Fast	\\
27.01.2006	&	53763.32655	&	330	&	1150	&	235	&	4.0098E+33	&	0.185	&	Fast	\\
27.01.2006	&	53763.35273	&	240	&	315	&	124	&	2.1100E+33	&	0.225	&	Slow	\\
02.02.2006	&	53769.28911	&	638	&	1471	&	245	&	4.1840E+33	&	0.040	&	Fast	\\
02.02.2006	&	53769.32073	&	1261	&	1802	&	270	&	4.6116E+33	&	0.118	&	Slow	\\
02.02.2006	&	53769.34732	&	495	&	920	&	114	&	1.9516E+33	&	0.118	&	Slow	\\
02.02.2006	&	53769.36786	&	855	&	864	&	126	&	2.1508E+33	&	0.010	&	Slow	\\
\enddata
\end{deluxetable}

\begin{deluxetable}{lcccccc}
\tabletypesize{\scriptsize}
\tablecolumns{7}
\tablewidth{0pc}
\tablecaption{The parameters derived from the \textit{OPEA} models, which were derived by using the Least-Squares Method.}
\\
\tablehead{
\colhead{Star}	&	\colhead{B-V}	&	\colhead{Plateau}	&	\colhead{$y_{0}$}	&	\colhead{$k$}	&	\colhead{Span Value}	&	\colhead{Half - Life}	\\
	&	\textbf{(mag)}	&	\colhead{($logP_{u}$)}	&	\colhead{($logP_{u}$)}	&	\colhead{(Tot. Duration)}	&	\colhead{($logP_{u}$)}	&	\colhead{(Tot. Duration)}}	\\
\\
\startdata 													
\textbf{V1005 Ori}	&	1.307	&	2.637 $\pm$ 0.074	&	0.428 $\pm$ 0.198	&	0.003063 $\pm$ 0.000623	&	2.209 $\pm$ 0.193	&	226.30	\\
\textbf{AD Leo}	&	1.498	&	2.527 $\pm$ 0.086	&	0.370 $\pm$ 0.054	&	0.002738 $\pm$ 0.000341	&	2.158 $\pm$ 0.089	&	253.10	\\
\textbf{V1054 Oph}	&	1.552	&	2.462 $\pm$ 0.105	&	0.385 $\pm$ 0.101	&	0.002305 $\pm$ 0.000435	&	2.077 $\pm$ 0.119	&	300.70	\\
\textbf{EV Lac} 	&	1.554	&	3.014 $\pm$ 0.084	&	0.698 $\pm$ 0.057	&	0.002404 $\pm$ 0.000250	&	2.316 $\pm$ 0.087	&	288.40	\\
\textbf{EQ Peg}	&	1.574	&	2.935 $\pm$ 0.091	&	0.859 $\pm$ 0.081	&	0.002074 $\pm$ 0.000272	&	2.076 $\pm$ 0.094	&	334.30	\\
\enddata
\end{deluxetable}

\begin{table}
\begin{center}
\caption{Using \textit{t-Test} analysis, computed mean values of equivalent durations for flares in the plateau phases are listed. In the first column, star names are given, while the mean values of equivalent durations are listed with their errors in the second one. In the last column, the standard deviations are given.\label{tbl-6}}
\begin{tabular}{lcc}
\tableline\tableline
\textbf{Star}	&	\textbf{Mean}	&	\textbf{Std. Deviation}	\\
\tableline 					
\textbf{V1005 Ori}	&	2.753 $\pm$ 0.051	&	0.196	\\
\textbf{AD Leo}	&	2.506 $\pm$  0.074	&	0.286	\\
\textbf{V1054 Oph}	&	2.269 $\pm$  0.078	&	0.257	\\
\textbf{EV Lac} 	&	3.165 $\pm$  0.101	&	0.285	\\
\textbf{EQ Peg}	&	3.021 $\pm$  0.095	&	0.232	\\
\tableline
\end{tabular}
\end{center}
\end{table}

\begin{table}
\begin{center}
\caption{The flare frequencies of stars for each observing season are listed. In the table, star name and observing seasons are given in first two columns. In the next column, the total observing durations ($\Sigma T_{t}$) are listed for each season. Then, the total numbers of flares obtained in a season ($\Sigma n_{f}$) are given. In $5^{th}$ column, the total equivalent durations obtained from all flares detected in that observing season ($\Sigma P_{u}$) are listed. In the last two columns, the flare frequencies ($N_{1}$ and $N_{2}$) are given.\label{tbl-7}}
\begin{tabular}{lcccccc}
\tableline\tableline
\textbf{Star}	&	\textbf{Season}	&	\textbf{$\Sigma T_{t}$ (hour)}	&	\textbf{$\Sigma n_{f}$}	&	\textbf{$\Sigma P_{u}$ (hour)}	&	\textbf{$N_{1}$ ($h^{-1}$)}	&	\textbf{$N_{2}$}	\\
\tableline 													
\textbf{V1005 Ori}	&	2004-2005	&	30.1403	&	10	&	0.529	&	0.332	&	0.018	\\
	                &	2005-2006	&	31.4011	&	31	&	2.100	&	0.987	&	0.067	\\
\tableline 													
\textbf{AD Leo}	&	2005	&	39.6908	&	39	&	3.396	&	0.983	&	0.086	\\
	            &	2006	&	40.5623	&	54	&	0.491	&	1.331	&	0.012	\\
	            &	2007	&	24.1147	&	17	&	0.168	    &	0.705	&	0.007	\\
\tableline 													
\textbf{V1054 Oph}	&	2004	&	42.6375	&	14	&	1.007	&	0.328	&	0.024	\\
                	&	2005	&	33.1622	&	26	&	0.296	&	0.784	&	0.009	\\
\tableline 													
\textbf{EV Lac} 	&	2004	&	54.1250	&	31	&	2.644	&	0.573	&	0.049	\\
	                &	2005	&	30.3108	&	32	&	0.940	&	1.056	&	0.031	\\
	                &	2006	&	48.7625	&	35	&	2.443	&	0.718	&	0.050	\\
\tableline 													
\textbf{EQ Peg}	&	2004	&	66.2792	&	38	&	1.364	&	0.573	&	0.021	\\
	            &	2005	&	37.4664	&	35	&	3.314	&	0.934	&	0.088	\\
\tableline
\end{tabular}
\end{center}
\end{table}

\end{document}